\documentclass[prd,onecolumn,notitlepage,showpacs,preprintnumbers,amsmath,amssymb,nofootinbib,APS,10pt,superscriptaddress]{revtex4-1}

\usepackage{dcolumn}
\usepackage{bm}
\usepackage[applemac]{inputenc}
\usepackage[spanish,english]{babel}
\usepackage{amsmath,amssymb,amsfonts,latexsym,cancel}
\usepackage{graphicx}
\usepackage{color}
\usepackage{soul}
\usepackage{ulem}
\usepackage{hyperref}
\usepackage{slashed}

\allowdisplaybreaks

\newcommand{\be}{\begin{equation}}
\newcommand{\ee}{\end{equation}}
\newcommand{\bea}{\begin{eqnarray}}
\newcommand{\eea}{\end{eqnarray}}

\begin{document}
\title{{\bf 
Note on the pragmatic mode-sum regularization method: Translational-splitting  in a cosmological background }}

\author{Pau Beltr\'an-Palau}\email{pau.beltran@uv.es}
\affiliation{Departamento de Fisica Teorica and IFIC, Centro Mixto Universidad de Valencia-CSIC. Facultad de Fisica, Universidad de Valencia, Burjassot-46100, Valencia, Spain.}
\author{Adri\'an del R\'io}\email{adrian.rio@uv.es}
\affiliation{Institute for Gravitation and the Cosmos \& Physics Department,The Pennsylvania State University, University Park, Pennsylvania 16802, USA}
\author{Sergi Nadal-Gisbert}\email{sergi.nadal@uv.es}
\affiliation{Departamento de Fisica Teorica and IFIC, Centro Mixto Universidad de Valencia-CSIC. Facultad de Fisica, Universidad de Valencia, Burjassot-46100, Valencia, Spain.}
\author{Jos\'e Navarro-Salas}\email{jnavarro@ific.uv.es}
\affiliation{Departamento de Fisica Teorica and IFIC, Centro Mixto Universidad de Valencia-CSIC. Facultad de Fisica, Universidad de Valencia, Burjassot-46100, Valencia, Spain.}

\begin{abstract}
The point-splitting  renormalization method offers a prescription to calculate finite expectation values of quadratic operators constructed from quantum fields in a general curved spacetime.  It has been recently shown  by Levi and Ori that when the background metric possesses an isometry, like stationary or spherically symmetric black holes, the method can be upgraded  into a pragmatic  procedure of renormalization that produces efficient numerical calculations. In this paper we  show that when the background enjoys three-dimensional spatial symmetries, like homogeneous expanding universes, the above pragmatic  regularization technique reduces to the well-established adiabatic regularization method.   


 \end{abstract}


\date{\today}
\maketitle



\section{Introduction}

Obtaining accurate theoretical predictions from quantum field theory  has become a topic of great interest nowadays for studies of the early universe and black holes. The naive calculation of physical observables  associated with a quantum field $\phi$, such as $\left< \phi^2(x)\right>$ or $\left<T_{\mu\nu}(x)\right>$, typically leads to  divergent sums or integrals of field modes, thereby requiring  the study of renormalization.
 While the systematics of renormalization in a general curved spacetime has been known for several decades now \cite{birrell-davies,Fulling,Wald94,parker-toms,DeWitt75}, the implementation of the standard prescription  to get specific results is still difficult to put in practice even for the most simple spacetime backgrounds. 
This is  because the regularization of ultraviolet divergences in a covariant way, and the construction of the subtraction terms, are based on the  point-splitting technique \cite{DeWitt65, christensen1976, christensen1978}, a purely analytical procedure that involves taking limits of points along geodesics. However, getting the field modes in a given spacetime background  requires solving complicated  differential equations, which can only be addressed numerically but in exceptional cases. A procedure to transform the covariant point-splitting technique into a numerically implementable method is thus almost mandatory if quantum field theory aims to produce results of practical interest for most gravitational scenarios. 

The numerical implementation of the point-splitting regularization method is however a nontrivial task, specially for black hole backgrounds. In a Schwarzschild  metric, the first important insight was introduced by Candelas in \cite{candelas1980}  by proposing an integral representation of the subtraction terms  of point-splitting,   allowing the possibility of subtracting the ultraviolet  divergences within the integral of field modes,  thereby yielding a formally finite result upon which the limit of points could be  taken in advance. However, the numerical implementation of these integrals was still a difficult task and this idea was not pursued further. 
An alternative way to address the problem was proposed in \cite{candelas-howard1984}, which did not involve the numerical evaluation of integrals, but which required an analytic WKB-type approximation of the field and a Wick rotation to analytically extend the metric to the Euclidean space.  
This method was successful in the Schwarzschild background and it was later extended for a general static and spherically symmetric metric in \cite{Andersonetal}.
%

Unfortunately, these analytical techniques are not available for  time-dependent backgrounds, as for instance in gravitational collapse, and thus this approach could not be extended to dynamical settings, that are of great interest in astrophysics. 
This problem recently motivated Levi and Ori \cite{levi-ori2015, levi-ori2016prl} to develop what they called the pragmatic mode-sum method of regularization, which  bypasses any   analytic approximation,  and that can be applied to any background metric as long as it displays an isometry.   Their approach recovers the first insight proposed by Candelas of finding integral representations of the point-splitting subtraction terms,  and proposes a successful method to implement numerically the integration over the field modes, based on the concept of generalized integrals. The technique has been proven to be  useful in computing numerically $\left< \phi^2(x)\right>$ and $\left<T_{\mu\nu}(x)\right>$ in 
black hole backgrounds  in different complementary ways \cite{levi2017b, levi-ori2016, levi2017, levi-ori2018, levi-ori2018b, ori2019, levi-ori2020}. 


The importance of this new approach is   that it can be applied to any metric provided that it has some symmetry.
For instance, in order to calculate $\langle \phi^2(x)\rangle$  one splits the points as $\langle \phi(x)\phi(x')\rangle$, which is well defined, and then, after subtracting the necessary DeWitt-Schwinger  term $G_{DS}(x,x')$, one takes the limit in which the two points $x, x'$ merge: $\langle \phi^2(x)\rangle=\lim_{x'\to x}(\langle \phi(x)\phi(x')\rangle- G_{DS}(x,x') )$. Following the proposal in 
 \cite{levi-ori2015}, in a stationary background  there is a preferred direction for which this splitting could be taken, which is the direction of the time-translational Killing vector field, i.e. $x = (t,r,\theta,\varphi)$ and $x'= (t+\epsilon, r, \theta,\varphi)$ in the usual Boyer-Lindquist coordinates. Furthermore,   the field $\phi(x)$ can be expanded in modes of well-defined frequency. 
Then, the point-splitting in the symmetric direction allow us to write the subtraction term $G_{DS}(x,x')$ as integrals in the field-mode frequencies, by Fourier-transforming  each term with respect to the splitting parameter $\epsilon$. 
The resulting expression for the difference $\langle \phi(x)\phi(x')\rangle- G_{DS}(x,x')$ is  an integral in frequencies which is formally finite in the limit $x'\to x$, so this limit can  be safely taken inside the integral. This is specially appropriate for numerical implementation, since only an integration is required to get the desired final result. It is important to stress that for the whole procedure to be well defined the  time-translational symmetry is fundamental. Similar reasonings can be applied with spherical symmetry (implemented via angular splitting), or only axial symmetry (implemented via azimuthal point-splitting) \cite{levi-ori2015,levi-ori2016prl,levi-ori2016,levi2017b,levi2017}.

So far the pragmatic mode-sum regularization method has been only applied  for stationary black hole spacetimes. Can one implement this procedure for other, possibly dynamical, symmetric spacetimes?
 Friedmann-Lemaitre-Robertson-Walker (FLRW) spacetimes, which are of interest in studies of cosmology, have three spatial Killing vector fields associated with spatial translations. Consequently, it is  natural  to use those symmetries to upgrade the point-splitting method and rewrite the subtraction term $G_{DS}(x,x')$ as an integral in modes of momentum $\vec{k}$ (i.e., the constants of motion associated with the spatial translation symmetries). 
 The goal of this work is to carry out this simple idea. As a result, we shall find that  the subtraction integrals match the expressions offered by the  method of adiabatic regularization \cite{parker-fulling}.

 The paper is organized in the following way. In Sec. \ref{SectionTemporalSplitting} we outline the main idea of the   pragmatic mode-sum regularization method. To this end we restrict the presentation to a stationary background and evaluate the renormalized two-point function $\langle \phi^2(x) \rangle$ by splitting the  points in the associated timelike direction.
 In Sec. \ref{SectionTranslationlSplitting} we extend the method to provide a numerically implementable formula for $\langle \phi^2(x) \rangle$ in a spatially flat FLRW spacetime using  the spatial translational symmetry of the metric.   
We end the section  generalizing the method by considering an arbitrary renormalization point $\mu$. In this work we follow the conventions in \cite{parker-toms}; in particular we use the metric signature  $(+,-,-,-)$. \\

\section{Pragmatic mode-sum regularization method in a stationary background: $t$-splitting}\label{SectionTemporalSplitting}
In this section we outline the idea underlying the pragmatic mode-sum regularization method introduced in \cite{levi-ori2015}, emphasizing those aspects that are relevant for our purposes.  The method takes advantage of the symmetries of the spacetime metric to rewrite the renormalization subtractions in the point-splitting method into a numerically efficient way. 
In order to illustrate the procedure, let us focus on the computation of the two-point function of a scalar field by exploiting the stationary symmetry.

Let $\phi(x)$ be a scalar  field of mass $m$ living in a  stationary spacetime of metric $g_{\mu\nu}$ that obeys the field equation  $(\Box + m^2 +\xi R) \phi =0$, where $\xi$ is the coupling constant to the scalar curvature $R$. To formulate a quantum description of this field, one must construct a Hilbert space of states.  As is well known, in a general curved spacetime there is no preferred prescription to do this. However, if the  spacetime is stationary one can define creation and annihilation operators, $A^{\dagger}_{\omega}$ and $A_{\omega}$, by decomposing the field operator into positive and negative frequency parts, 
\be \phi(x)= \int_m^\infty d\omega  \left[  A_{\omega }f_{\omega }(x) +  A^{\dagger}_{\omega }f^*_{\omega }(x)\right] \ ,  \label{fielddecomposition}
\ee
and define the vacuum state using the annihilation operators.  The notion of  field modes $f_{\omega}$, $f^*_{\omega}$ of positive and negative frequency $\omega$ can be introduced in a natural way by the conditions $\mathcal L_K f_{\omega }=-i \omega f_{\omega }$, $\mathcal L_K f^*_{\omega }=i \omega f^*_{\omega }$, where $K$  is the infinitesimal generator of the isometry (i.e. the Killing vector field) \cite{DeWitt75}.  Then, using this set of field modes there is a preferred prescription to decompose the field operator as above \cite{AshtekarMagnon75}. We have omitted   the additional quantum numbers required to specify a basis of  modes, since they do not play any fundamental role in the following discussion.
%
Choosing now a natural coordinate system $\{t, x^k\}$ such that $K=\partial/\partial t$, the above conditions imply
\be
f_{\omega }(x)=e^{-i\omega t}\psi_{\omega }(\vec x) \, , \label{modefunctions}
\ee
where $\vec x$ is a shorthand for the three spatial coordinates $x^k$. The determination of the spatial functions $\psi_{\omega }(\vec x)$ is achieved by solving numerically the Klein-Gordon equation. 

The naive calculation of  $\langle\phi^2(x)\rangle$ will produce a divergent expression, as expected, so a renormalization method is needed at this point. 
The DeWitt-Schwinger point-splitting method consists in taking the product of the field operator at two separated points $x, x'$;  subtracting to this two-point function the corresponding asymptotic DeWitt-Schwinger proper-time expansion up to the last divergent term; and finally taking the coincident limit $x\to x'$ along a geodesic that connects the two points.
The result of this procedure is what defines the renormalized two-point function: 
\be \label{TPF}
\left\langle\phi^{2}(x)\right\rangle_{\mathrm{ren}}=\lim _{x^{\prime} \rightarrow x}\left[ \left\langle \{\phi(x), \phi\left(x^{\prime}\right)\}\right\rangle-G^{(1)}_{DS}\left(x, x^{\prime}\right)\right] \, ,
\ee
where $\{\phi(x), \phi(x')\} \equiv \frac{1}{2}[\phi(x)\phi(x')+\phi(x')\phi(x)]$, and 
$G_{DS}^{(1)}\left(x, x^{\prime}\right)$ is the symmetric part of the DeWitt-Schwinger subtraction term (Hadamard function). 
Following \cite{christensen1976}, it is possible to obtain an expansion for the symmetric two-point function in terms of covariant quantities evaluated at $x$ and the geodesic distance between $x$ and $x'$. 
Including only the relevant terms needed in the calculation of the renormalized two-point function (i.e. up to second-order derivatives of the metric), the subtraction term yields
\be \label{GDS}
  G^{(1)}_{DS} \left(x, x^{\prime}\right)= \frac{1}{8 \pi^{2}}\left[- \frac{1}{\sigma}+(m^{2}+(\xi-1 / 6) R)\left(\gamma+\frac{1}{2} \log \left(\frac{m^{2} |\sigma|}{2}\right)\right)
-\frac{m^{2}}{2}+\frac{1}{12} R_{\alpha \beta}\frac{\sigma^{; \alpha} \sigma^{; \beta}}{\sigma}\right]\, ,
\ee
where $R$ is the scalar curvature and  $R_{\alpha\beta}$ is the Ricci tensor. $\gamma$ is the Euler constant and  $\sigma(x,x')=\frac12\tau(x,x')^2$,  $\tau(x,x')$ being the proper distance along the geodesic connecting $x$ to $x'$ (for sufficiently close points this geodesic is unique \cite{oneil}).
The expansion (\ref{GDS}) contains all the divergences of the two-point function.
As a remark, expression (\ref{GDS}) finds its origin in the integral expression for the Feynman Green function
\be\label{GreenFuncProperTime1}
G_{DS}(x,x') = \frac{\Delta^{1/2}(x,x')}{(4\pi)^2} \int^{\infty}_{0} \frac{d s}{(is)^2}e^{-i(m^2 s + \frac{\sigma}{2 s})} \sum^{\infty}_{n=0}a_n(x,x') (i s)^n  \ ,
\ee
where $a_n(x,x')$ are the DeWitt coefficients, which can be solved recursively  from the field equations using the input $a_0(x,x')=1$ \cite{DeWitt75}, and $\Delta(x,x')$ is the  Van Vleck-Morette determinant defined as \be \Delta(x,x')=- |g(x)|^{-1/2} det\left[ -\partial_{\mu}\partial_{\nu'}\sigma(x,x')\right]|g(x')|^{-1/2} \ . \ee The above expression can be written in terms of  Hankel functions  which, after expanding in an asymptotic series, give rise to \eqref{GDS}. 
As we will see in Sec. \ref{SectionGeneralizedRegu}, Eq. \eqref{GreenFuncProperTime1} will be the key starting point to generalize the subtraction terms in order to deal with the infrared divergence when $m \rightarrow 0$  (by means of the introduction of an arbitrary renormalization point $\mu$). To implement the renormalization prescription with the pragmatic mode-sum regularization we just need \eqref{GDS}, so we will forget about \eqref{GreenFuncProperTime1} for the moment.


At this point it becomes evident that, if the mode functions in (\ref{modefunctions}) are to be solved numerically, the explicit calculation of (\ref{TPF}) with (\ref{GDS}) using numerical methods is far from obvious. Here is where the pragmatic mode-sum method comes into play. 
Following \cite{levi-ori2015}, we have to split the points $x$ and $x'$ in the direction associated with the symmetry, i.e., such that the metric has the same value in both points. Choosing  $x=(t,\vec x)$ and $x'=(t+\epsilon, \vec x)$ with $\epsilon>0$ an infinitesimal parameter, the mode expansion of the symmetric two-point function formally reads
\be \label{TPFdiv}
\langle \{\phi(x),\phi(x')\} \rangle =\int_m^\infty d\omega \cos\left(\omega\epsilon \right) | \psi_{\omega }(\vec x)|^2\ . 
\ee
On the other hand, 
expanding $\sigma$ in a Taylor series around $\epsilon=0$  one finds that the general form of $G_{DS}^{(1)}(x,x')$ has the form
\be
 G^{(1)}_{DS}(x,x')=a( \vec x)\frac{1}{\epsilon^2}+c( \vec x)\left(\log\left(m\epsilon \right)+\gamma \right)+d(\vec x)+\mathcal{O}(\epsilon)\, ,
\ee
where $a(\vec x), c(\vec x)$, and $d(\vec x)$ are real functionals  of the metric.
 The key point now is to express the $\epsilon$-dependent terms  as  integrals in $\omega$ by using the following integral transforms:
\bea
&&\int_m^\infty d\omega \ \omega  \cos(\omega \epsilon )=-\frac{1}{\epsilon^2} - \frac{m^2}{2} +\mathcal{O}(\epsilon)\, ,\\
&& \int_m^\infty\frac{d\omega}{\omega+ m}\cos(\omega\epsilon)=-(\log( m \epsilon)  +\gamma) -\log2 +\mathcal{O}(\epsilon) \ .
\eea
These integrals have to be understood as generalized integrals in the distributional sense. Inserting all these expressions in \eqref{TPF} one finds the following expression for the renormalized two-point function:
\be
\langle\phi^2(x)\rangle_{ren}=\lim_{\epsilon\to0}\int_m^\infty d\omega \left( |\psi_{\omega }( \vec x)|^2+a( \vec x)\omega+c( \vec x)\frac{1}{\omega+m}\right)\cos\left( \omega \epsilon\right) -\bar{d}( \vec x)\, ,
\ee
where $\bar{d}( \vec x)= d( \vec x) -a( \vec x)\frac{m^2}{2} - c(\vec{x}) \log 2$. The integral is now expected to be convergent since the original point-splitting subtraction terms have been designed to cancel the divergences of the two-point function.
Then, the limit and the integration can be interchanged and we finally obtain the following result for the renormalized two-point function:
\be \label{TPFren}
\langle\phi^2(x)\rangle_{ren}=\int_m^\infty d\omega \left( |\psi_{\omega }(\vec x)|^2+a(\vec x)\omega+c(\vec x)\frac{1}{\omega+m}\right)-\bar{d}(\vec x)\, .
\ee

Thus, with these manipulations the quantity $\langle\phi^2(x)\rangle_{ren}$ can, at least in principle, be computed using ordinary numerical techniques.\footnote{Notice though that the existence of an isometry was fundamental. Had the coordinate $t$ failed to be associated with a Killing vector field, the above procedure could not have been carried out.}
In practice, however, there is one last issue that must be addressed, at least in some cases. To give an example, in a Schwarzschild background  and for a massless field $m=0$ one gets \cite{levi-ori2015} $a=-(4\pi^2(1-2M/r))^{-1}$, $c=0$, $d=M^2(48\pi^2r^4(1-2M/r))^{-1}$, which agrees with the result originally introduced in \cite{candelas1980}. The point is that, when trying to implement the above integration numerically, one finds that it fails to converge. This is because when performing the integration between 0 and $\omega$, increasing oscillations in $\omega$ appear. As pointed out in \cite{levi-ori2015}, the origin of these oscillations comes from the fact that black holes admit null geodesics connecting $x$ and $x'$, i.e., geodesics that start at some spatial point and after making one or several round trips around the black hole return to the same point with a delay  time given by $\epsilon$. At the values of $\epsilon$ corresponding to these geodesics the term $\left\langle\phi(x) \phi\left(x^{\prime}\right)\right\rangle$ presents singularities, which in Fourier domain is equivalent to oscillations in $|\psi_{\omega}(\vec x)|^2$ [see  Eq. \eqref{TPFdiv}]. The wavelengths of the oscillations are related to the values of $\epsilon$ of these geodesics, that can be obtained through a straightforward analysis of the geodesic equation in Schwarzschild spacetime. To solve the issue of the divergent integration, one can apply a ``self-cancellation'' numerical method,  explained in detail in \cite{levi-ori2015}, in order to cancel the oscillations and obtain the physical finite value of the renormalized two-point function. Fortunately, in the case we will study in this work these kinds of geodesics do not exist, so there will not be any convergence problem, and then the self-cancellation method will not be necessary.

As a final remark, for massless fields the term $\log \left(m^2 |\sigma|/2\right)$ in (\ref{GDS}) is ill defined. This infrared problem  is usually bypassed by replacing the mass by a new arbitrary parameter $\mu$ in the logarithm. Therefore, in the massless case \eqref{TPFren} actually reads 
\be \label{TPFrenmasless}
\langle\phi^2(x)\rangle_{ren}=\int_0^\infty d\omega \left( |\psi_{\omega }(\vec x)|^2+a(\vec x)\omega+c(\vec x)\frac{1}{\omega+\mu}\right)-d(\vec x)\, .
\ee
 We will reconsider this point later on, specially in the quantization of the field in the FLRW background.\\

\section{Pragmatic mode-sum regularization method in a FLRW background: Translational-splitting}\label{SectionTranslationlSplitting}

As pointed out in the Introduction, our aim is to extend the pragmatic mode-sum method to a cosmological setting. We shall work out  the case of a scalar field in  a FLRW spacetime, and  consider a spatially flat universe with metric  $ds^2=dt^2- a^2(t)d\vec{x}^2$, for simplicity. 
This spacetime is dynamical and $t$-splitting is no longer useful. On the contrary, since the background we are considering now is spatially homogeneous, it is natural to use the translational symmetry when applying the point-splitting prescription. 

Given the spatial homogeneity of the spacetime background, the field  operator $\phi$ can now be naturally expanded  in the form
\be \phi(x)= \int d^3k \left[  A_{\vec{k}}f_{\vec{k}}(x) +  A^{\dagger}_{\vec{k}}f^*_{\vec{k}}(x)\right] \ , 
\ee
where, again, $A_{\vec{k}}$ and $A^\dagger_{\vec{k}}$ are  annihilation and creation operators satisfying canonical commutation relations, and $f_{\vec k}(x)$ 
denote a complete orthonormal family of solutions to the field equation  satisfying $\mathcal L_{K^j} f_{\vec{k}} = i k^j f_{\vec{k}}$, where $\{K^j\}_{j=1,2,3}$ denote the three Killing vector fields associated with spatial translations. Thus, in a canonical coordinate chart $\{t, \vec x\}$ where $K^j=\partial/\partial x^j$ the field modes take the general form
\be\label{ModesFLRW}
f_{\vec{k}}(x)=\frac{e^{i\vec{k}\cdot \vec{x}  }    }{\sqrt{2(2\pi)^3a(t)^3}}h_k(t)  \ . 
\ee
 These modes are assumed to obey the normalization condition with respect to the conserved Klein-Gordon product $(f_{\vec k}, f_{\vec k'}) = \delta^3(\vec k - \vec k')$, $(f_{\vec k}, f^*_{\vec k'}) = 0$.
This condition translates into a Wronskian-type condition for the modes: $h_k^*\dot h_k - \dot h_k^{*} h_k = -2i$, where the dot means derivative with respect to time $t$.  The complete specification of the modes usually requires assuming boundary conditions at early times. This is however not relevant for renormalization. 

Let us now proceed to the regularization of the two-point function via the point-splitting method. As explained in the previous section the renormalized two-point function is defined as 
\be \label{TPFflrw}
\left\langle\phi^{2}(x)\right\rangle_{\mathrm{ren}}=\lim _{x^{\prime} \rightarrow x}\left[\left\langle\{\phi(x), \phi\left(x^{\prime}\right)\}\right\rangle-G_{\mathrm{DS}}^{(1)}\left(x, x^{\prime}\right)\right]
\ , \ee
where the DeWitt-Schwinger subtraction term is given by (\ref{GDS}), 
 now with $R=6(\frac{\ddot a}{a}+\frac{\dot a^2}{a^2}), R_{00}=3\frac{\ddot a}{a}, R_{ii}=-a^2(\frac{\ddot a}{a}+\frac{2\dot a^2}{a^2})$ for a FLRW metric.

As illustrated in the previous section the point-splitting regularization scheme becomes particularly useful when  we evaluate the two-point function in two points where the metric has the same value. Therefore, taking advantage of the translational symmetry of the FLRW  spacetime, we consider equal-time points $x\equiv (t,\vec x)$ and $x' \equiv (t,\vec x+\vec \epsilon)$ to do the splitting. Then, the equal-time two-point function reads
\be \langle \{\phi(x), \phi (x')\} \rangle = \frac{1}{2(2\pi a(t))^3}\int d^3 k |h_k(t)|^2 \cos{(\vec{k} \cdot \vec{\epsilon})} \ . \ee 
One can now express the cosine in terms of exponentials and easily perform angular integration to reduce it to 
\be \langle \{\phi(x),\phi(x')\}\rangle = \frac{1}{4 \pi^2 a(t)^3}\int_0^\infty dk k^2 |h_k(t)|^2 \frac{\sin k \epsilon}{k\epsilon} \ ,\ee 
where $k=|\vec k|$ and $\epsilon= |\vec \epsilon|$. To achieve an efficient numerical method of renormalization, the goal now is to rewrite the DeWitt-Schwinger term $G^{(1)}_{DS}(x,x')$ in (\ref{GDS}) as an integral in momentum space so that it can be fitted with the previous expression for the field modes. As in the previous section, we have to  evaluate  $\sigma$. To this end, it is useful to use the Riemann normal coordinates $y^{\mu}$ with origin at $x$. In these coordinates we have $\sigma(x,x')=\frac12 y_\mu y^\mu$. Following \cite{brewin2009}, $y^{\mu}$ can be expanded in terms of $\Delta x^{\mu}= x'^{\mu}-x^{\mu}$, and we  obtain (for the first orders)
\be
\sigma=\frac{1}{2}\Delta t^{2}-\frac{a^{2}}{2} \Delta \vec{x}^{2}-\frac{a \dot{a}}{2} \Delta \vec{x}^{2} \Delta t -\frac{a \ddot a}{6}  \Delta \vec{x}^{2} \Delta t^{2}-\frac{a^{2}\dot{a}^{2}}{24}\Delta \vec{x}^{4}+\cdots \label{sigmaAll}
\ . \ee
 Therefore in our case we can write
\be\label{SigmaExpansion}
\sigma=-\frac{a^2}{2}\epsilon^2-\frac{a^2\dot a^2}{24}\epsilon^4 -\frac{ \left(a^2 \dot{a}^4+3 a^3
   \dot{a}^2 \ddot{a}\right)}{720} \epsilon ^6+\mathcal{O}\left(\epsilon^8\right)\, .
\ee
The terms involving $\sigma$ in (\ref{GDS}) can be now expanded as follows [note that $\sigma^{;\alpha}$ must be calculated from (\ref{sigmaAll})]
\bea
&&\frac{1}{\sigma}  =  -\frac{2}{a^2 \epsilon^2}+\frac{\dot a^2}{6a^2}+\mathcal{O}(\epsilon^2) \ . \label{I1}\\
&&R_{\alpha \beta} \frac{\sigma^{; \alpha} \sigma^{; \beta}}{\sigma}
=2\frac{\ddot a}{a}+4\frac{\dot a^2}{a^2}+\mathcal{O}(\epsilon^2) \ .
\eea
Introducing these results in (\ref{GDS}) we get
\be \label{GDSflrw}
G_{\mathrm{DS}}^{(1)}(x,x') = \frac{1}{4\pi^2} \left(\frac{1}{a^2 \epsilon^2} + \frac12 (m^{2}+(\xi-1 / 6) R)\left(\gamma+ \log \left(\frac{ma}{2}\epsilon\right) \right)
-\frac{m^{2}}{4}+\frac{R}{72}\right) + \mathcal{O}(\epsilon) \ ,
\ee
which turns out to be a function depending on $\vec \epsilon$ only through its modulus $\epsilon$ (this is due to the underlying isotropy of the FLRW metric). Now we have to rewrite the potential divergences of this expression as $\epsilon\to 0$ in terms of one-dimensional integrals in momentum space involving $\frac{\sin k \epsilon}{k\epsilon}$. To this end we consider the following integral transforms, which have to be understood as generalized integrals: 
\bea
&&\int_0^\infty dk k \frac{\sin k \epsilon}{k\epsilon} = \frac{1}{\epsilon^2}\, ,\label{MomentumIntegral1}\\
&&\int_0^\infty dk \frac{k^2}{\omega^3} \frac{\sin k \epsilon}{k\epsilon} = -a^3\left(\gamma+ \log \left(\frac{ma}{2}\epsilon\right)\right)+ \mathcal{O}(\epsilon)\, ,\label{MomentumIntegral2}
\eea
where $\omega(t)= \sqrt{k^2/a^2(t) + m^2}$. Substituting in \eqref{GDSflrw} the terms depending on $\epsilon$ by these integrals we get
\be
G_{\mathrm{DS}}^{(1)}\left(x,x' \right) =
\frac{1}{4 \pi^2 a^3}  \int_0^{\infty} d k \frac{\sin (k\epsilon)}{k\epsilon}  \left[ ka-\frac{k^2m^2}{2\omega^3} + \frac{k^2(\frac{1}{6} -\xi)R}{2\omega^3}\right] -\frac{m^2}{16\pi^2}+ \frac{R }{288 \pi^2} + \mathcal{O}(\epsilon)\, .
\ee

Using now  the identity 
\be\label{MassInTermsOfIntegralk}
\frac{1}{4 \pi^2 a^3}\int_0^{\infty} d k \frac{\sin(k \epsilon)}{k\epsilon} \left[ ka-\frac{k^2m^2}{2\omega^3} -\frac{k^2}{\omega}\right]=\frac{m^2}{16\pi^2} + \mathcal{O}(\epsilon)\, ,
\ee
we can simplify the expression of $G_{DS}^{(1)}$. Introducing it into \eqref{TPFflrw} we can finally write
\be \label{TPFfinalepsilon}
\langle\phi^2\rangle_{ren}= \lim_{\epsilon \to 0} \frac{1}{4 \pi^2 a^3}  \int_0^{\infty} d k k^2 \frac{\sin k\epsilon}{k\epsilon} \,  \left[|h_k|^2- \frac{1}{\omega} - \frac{(\frac{1}{6} -\xi)R}{2\omega^3}\right] - \frac{R}{288 \pi^2} \, .
\ee
The sum of terms inside the parentheses has no ultraviolet divergences even for $\epsilon= 0$, so we can interchange the integral and the limit $\epsilon \to 0$ to find
\be \label{TPFfinal}
\langle\phi^2\rangle_{ren}=  \frac{1}{4 \pi^2 a^3}  \int_0^{\infty} d k k^2  \,  \left[|h_k|^2- \frac{1}{\omega} - \frac{(\frac{1}{6} -\xi)R}{2\omega^3}\right] - \frac{R}{288 \pi^2} \, .
\ee
Note that the above expression can be naturally expressed in terms of a three-dimensional integral in the $\vec k$ modes associated with the three-dimensional translation symmetry
\be \label{TPFfinal3k}
\langle\phi^2\rangle_{ren}=\frac{1}{2(2 \pi a)^3}  \int d^3k \,  \left[|h_k|^2- \frac{1}{\omega} - \frac{(\frac{1}{6} -\xi)R}{2\omega^3}\right] - \frac{R}{288 \pi^2} \, .
\ee
This result agrees exactly with the renormalized two-point function obtained by using the so-called adiabatic regularization method developed by Parker and Fulling in the early 1970s for cosmological backgrounds and scalar fields \cite{parker-fulling} (see \cite{delrio-navarro2015} for spin-$1/2$ fields). One can check the result with Eq. \eqref{GeneralizedFinalCountertermAD} in Appendix A. 
As explained before, the problem of frequency oscillations in the Schwarzschild black hole context,  pointed out in \cite{levi-ori2015}, does not emerge for FLRW metrics. Therefore, after solving numerically the Klein-Gordon equation for the modes $h_{k}(t)$, ordinary numerical integration techniques can be applied directly to calculate the final expression (\ref{TPFfinal}).

It is also straightforward to see that the result can be extended to the renormalized stress-energy tensor in a FLRW background. To this end one needs a more complete form of the DeWitt-Schwinger expansion of the two-point function \cite{christensen1976,bernard-folacci1986} 
\bea\label{christensen3.17}
G^{(1)}_{DS}\left(x, x^{\prime}\right)&=&\frac{\Delta^{1 / 2}}{8 \pi^{2}}\left\{ -\frac{1}{\sigma}+m^{2}\left(\gamma+\frac{1}{2} \ln \left|\frac{1}{2} m^{2} \sigma\right|\right)\left(1-\frac{1}{4} m^{2} \sigma\right)-\frac{1}{2} m^{2}+\frac{5}{16} m^{4} \sigma \right. \\
&&\left.-a_{1}\left[\left(\gamma+\frac{1}{2} \ln \left|\frac{1}{2} m^{2} \sigma\right|\right)\left(1-\frac{1}{2} m^{2} \sigma\right)+\frac{1}{2} m^{2} \sigma\right] -\frac12 a_{2} \sigma\left[\gamma+\frac{1}{2} \ln \left|\frac{1}{2} m^{2} \sigma\right|-\frac12\right]+\frac{a_2}{2 m^{2}}\right\} \ ,\nonumber
\eea
 where $a_1$ and $a_2$ are the first DeWitt coefficients. The renormalized vacuum expectation value of the stress-energy tensor can be obtained by acting with a nonlocal operator to the renormalized symmetric part of the two-point function 
\be\label{Tmunu}
\langle T_{\mu\nu}\rangle = \lim_{x \rightarrow x'}\mathcal{D}_{\mu\nu}(x,x')\left[\left\langle \{\phi(x), \phi\left(x^{\prime}\right)\}\right\rangle-G^{(1)}_{DS}(x,x') \right] \ .
\ee
This differential operator $\mathcal{D}_{\mu\nu}(x,x')$ contains different quadratic terms of covariant derivatives 
\cite{bernard-folacci1986,christensen1976}; therefore, we need to expand \eqref{christensen3.17} up to and including the order $\mathcal{O}(\epsilon^2)$ because terms proportional to $\epsilon^2$ can give rise to finite terms in $\langle T_{\mu\nu}\rangle $. 
Proceeding as before and expanding \eqref{christensen3.17} 
to order $\mathcal{O}(\epsilon^2)$ we arrive at the expression (\ref{GDS4eps}) (see Appendix B for details) which contains terms with four derivatives of the metric. 
This expression agrees with the subtraction terms of the two-point function obtained by adiabatic regularization at fourth adiabatic order (\ref{adiabaticST2sp}). 
Therefore, the pragmatic form of the subtraction terms for the stress-energy tensor, when the translational symmetry is considered, reduces to the renormalization terms of adiabatic regularization. 
The explicit formulas of interest required to do the direct numerical implementation can be seen, for instance, in \cite{anderson-parker1987}. These results explain the great versatility of the adiabatic method with numerical calculations \cite{Birrell78, hu-parker1977, anderson1985, anderson-eaker1999}.\\

\subsection{Massless case and the renormalization scale $\mu$} 
\label{SectionGeneralizedRegu}

For massless fields expression (\ref{GDS}) is ill defined due to a logarithmic divergence. The usual approach to bypass this infrared divergence is to introduce an upper cutoff in the proper-time integral (\ref{GreenFuncProperTime1}) \cite{DeWitt75}, or to replace $m^2$ by an arbitrary mass scale $\mu^2$ in the problematic logarithmic term. Here  we will follow an alternative strategy based on \cite{ferreiro-navarro2020} 
that consists in replacing $m^2$ by $m^2 + \mu^2$ in the exponent of the  DeWitt-Schwinger integral form  \eqref{GreenFuncProperTime1}.
The  advantage of this approach is that it leads to a natural decoupling mechanism of heavy massive fields. 
Following this idea, we  have
\be\label{GreenFuncProperTimemu}
G_{DS}(x,x')= \frac{\Delta^{1/2}}{(4\pi)^2} \int^{\infty}_{0} \frac{d s}{(i s)^2}e^{-i\left((m^2 + \mu^2)s + \frac{\sigma}{2 s}\right)} \sum^{\infty}_{n=0}\bar{a}_n (i s)^n \ .
\ee
In order to be consistent with \eqref{GreenFuncProperTime1}, the first DeWitt coefficients need to be modified  in the following way: $\bar{a}_0(x,x')=1, \ \bar{a}_1(x,x')=a_1(x,x')+\mu^2,\ \bar{a}_2(x,x')=a_2(x,x')+ a_1(x,x')\mu^2+\frac12\mu^4$. Now one can proceed as in the case in which $\mu=0$. We can write \eqref{GreenFuncProperTimemu} in terms of  Hankel functions  and later expand them in  asymptotic series to finally get the following expression for the subtraction term of the two-point function in the point-splitting renormalization method,
\be \label{GeneralizedCountertermMu}
  G^{(1)}_{DS} \left(x, x^{\prime}\right)= \frac{1}{8 \pi^{2}}\left[- \frac{1}{\sigma}+(m^{2}+(\xi-1 / 6) R)\left(\gamma+\frac{1}{2} \log \left(\frac{m^{2}+\mu^2 }{2}|\sigma|\right)\right)
-\frac{m^{2}+\mu^2}{2}+\frac{1}{12} R_{\alpha \beta}\frac{\sigma^{; \alpha} \sigma^{; \beta}}{\sigma}\right]\, .
\ee
Note that  the  parameter $\mu^2$  appears nontrivially in this expression. Not only does it appear in the logarithmic term but it also emerges in the constant term, and not in the usual combination $m^{2}+(\xi-1 / 6) R$ multiplying the logarithm. This effect is responsible of the decoupling of heavy particles in the computations of the renormalized energy-momentum tensor \cite{ferreiro-navarro2020}. 

Considering a FLRW spatially flat spacetime we can write the generalized subtraction term with integrals in modes of $k$  by using again the translational symmetry. To do so, we just have to replace $m^2$ by $m_{\rm eff}^2 = m^2 + \mu^2 $ in  \eqref{MomentumIntegral2} and using the integral representations of the divergent terms [Eqs. \eqref{MomentumIntegral1} and \eqref{MomentumIntegral2})] in \eqref{GeneralizedCountertermMu}  we get
\be
G_{\rm DS}^{(1)}(x,x') = \frac{1}{4 \pi^2 a^3}  \int_0^{\infty} d k \frac{\sin(k\epsilon)}{k\epsilon} \left[ ka-\frac{k^2m^2}{2\omega_{\rm eff}^3} + \frac{k^2(\frac{1}{6} -\xi)R}{2\omega_{\rm eff}^3}\right] -\frac{m_{\rm eff}^2}{16\pi^2}+ \frac{R}{288 \pi^2}+ \mathcal{O}\left(\epsilon\right)\, ,
\ee
where $\omega_{\rm eff}^2 = \frac{k^2}{a^2} + m^2 + \mu^2$. If we consider the identity \eqref{MassInTermsOfIntegralk} with $m^2$ replaced by $m_{\rm eff}^2$ we can rewrite the expression above as follows:
\be\label{GeneralizedFinalCountertermPMS}
G_{\rm DS}^{(1)}(x,x') =
\frac{1}{4 \pi^2 a^3}  \int_0^{\infty} d k k^2\frac{\sin(k\epsilon)}{k\epsilon} \,  \left[ \frac{1}{\omega_{\rm eff}} + \frac{(\frac{1}{6} -\xi)R}{2\omega_{\rm eff}^3} + \frac{\mu^2}{2\omega_{\rm eff}^3}\right] + \frac{R}{288 \pi^2}+ \mathcal{O}\left(\epsilon\right) \, .
\ee
This is the generalized subtraction term for the two-point function written as an integral in modes of the momentum $k$. Note that a new term proportional to $\mu^2$ appears. 
This result agrees with adiabatic regularization when we introduce the arbitrary parameter $\mu$ requiring the same conditions (see Appendix A for details). \\

\section{Conclusions}

 In this work we have applied the pragmatic mode-sum regularization method proposed by Levi and Ori to study the numerical implementability of renormalization for quantum fields in FLRW spacetime backgrounds. This was possible thanks to the isometry under spatial translations of the underlying metric. 
 The results obtained are in agreement with the well-known prescription of adiabatic regularization, developed by Parker and Fulling in the early 1970s. Adiabatic regularization can now be understood as the natural renormalization procedure that emerges when the point-splitting technique is applied using the spatial isometries of the FLRW metric.\\

\section{ Acknowledgments}\nonumber
This work has been supported by research Grants 
No.\  FIS2017-84440-C2-1-P; No. \ FIS2017-91161-EXP and the project PROMETEO/2020/079 (Generalitat Valenciana).  A.d.R. acknowledges support under NSF Grant No. PHY-1806356 and the Eberly Chair funds of Penn State. P. B. is supported by the Ministerio de Ciencia, Innovaci\'on y Universidades, Ph.D. fellowship No. FPU17/03712. S. N. is supported by the Universidad de Valencia, within the Atracci\'o de Talent Ph.D fellowship No. UV-INV- 506 PREDOC19F1-1005367.

\appendix
\section{Adiabatic regularization with an arbitrary $\mu$}\label{AppendixAdiabatic}
In this appendix we provide a very concise presentation of the adiabatic regularization method. Furthermore we also introduce the adiabatic procedure in a generalized way so as to account for the introduction of a renormalization scale $\mu$. 
It was first sketched  in \cite{ferreiro-navarro2019} by replacing $m^2$ by $\mu^2$ in the zeroth adiabatic order. However, following \cite{ferreiro-navarro2020} a  better, and physically motivated procedure, is to replace $m^2 \rightarrow m^2 + \mu^2$. The underlying reason is to guarantee the decoupling of heavy massive fields. 


Adiabatic renormalization is based on a generalized WKB-type asymptotic expansion of the modes \eqref{ModesFLRW} according to the ansatz
\be \label{kgansatz}  h_k(t) \sim \frac{1}{\sqrt{W_k(t)}}e^{-i\int^t W_k(t')dt'}  \ , \ee
which guarantees  the Wronskian condition $h_k \dot{h}^{*}_k - h_k^{*}\dot{h}_{k}= -2 i$.
One then expands $W_k$ in an adiabatic series, in which each contribution is determined by the number of time derivatives of the expansion factor $a(t)$
\be \label{adiabatic}  W_k(t) = \omega^{(0)}(t) + \omega^{(2)}(t)+ \omega^{(4 )}(t) + ... \ , \ee
where the leading term $\omega^{(0)}(t)\equiv \omega(t)= \sqrt{k^2/a^2(t) + m^2}$ is the usual physical frequency. Higher order contributions can be univocally obtained by iteration (for details, see \cite{parker-toms}),
which come from introducing (\ref{kgansatz}) into the equation of motion for the modes. The adiabatic expansion of the modes can be easily translated  to an expansion of the two-point function $\langle \phi (x) \phi (x') \rangle \equiv G(x, x')$ at coincidence $x=x'$:
\be G_{Ad}(x,x) = \frac{1}{2(2\pi)^3 a^3} \int d^3 k\,  [ \omega^{-1} + (W^{-1})^{(2)} +  (W^{-1})^{(4)} + ... ] \ . \ee
As remarked above, the  expansion must be truncated to the minimal adiabatic order necessary to cancel all ultraviolet divergences that appear in the formal expression of the vacuum expectation value  that one wishes to compute. The calculation of the renormalized variance $\langle \phi^2 \rangle$ requires only second adiabatic order. 

The above process can be repeated now by replacing $m^2$ by $m^2 + \mu^2$ in the zeroth adiabatic order $\omega$. Therefore the expansion for $W_k$ depends now on $\mu$
\be \label{adiabatic}  W_k(t) = \omega_{\rm eff}^{(0)}(t,\mu) + \omega_{\rm eff}^{(2)}(t,\mu)+ \omega_{\rm eff}^{(4 )}(t,\mu) + ... \ , \ee
where the leading term is $\omega_{\rm eff}^{(0)}(t)\equiv \omega_{\rm eff}(t)= \sqrt{k^2/a^2(t) + m^2 + \mu^2}$. The higher orders are univocally recalculated. For the second order $\omega_{\rm eff}^{(2)}(t,\mu)$ which is enough to renormalize the two-point function we have
\be
\omega_{\rm eff}^{(2)}=\frac{5 k^4 \dot{a}^2}{8 \omega_{\rm eff} ^{5} a^6}-\frac{3 k^2 \dot{a}^2}{4 \omega_{\rm eff}
   ^{3} a^4}+\frac{k^2 a}{4 \omega_{\rm eff} ^{3} a^3}+\frac{3 \xi 
   \dot{a}^2}{\omega_{\rm eff}  a^2}+\frac{3 \xi  \ddot{a}}{\omega_{\rm eff}  a}-\frac{3 \dot{a}^2}{8
   \omega_{\rm eff}  a^2}-\frac{3 \ddot{a}}{4 \omega_{\rm eff}  a}-\frac{\mu ^2}{2 \omega_{\rm eff}
   } \ .
\ee
The new terms proportional to $\mu^2$, serve to remove the divergences, in accordance with the new definition of $\omega_{\rm eff}^{(0)}(t,\mu)$ while maintaining locality and general covariance. Note that $\mu^2$ should be
regarded as a parameter of adiabatic order 2.

Therefore, the subtraction term for the two-point function is given by 
\be ^{(2)}G_{Ad}(x,x) = \frac{1}{2(2\pi)^3 a^3} \int d^3 k\,  \left[ \frac{1}{\omega_{\rm eff}} - \frac{\omega_{\rm eff}^{(2)}}{\omega_{\rm eff}^2} \right] \ . \ee
After a little bit of algebra the terms in the integral can be written like
\be ^{(2)}G_{Ad}(x,x) = \frac{1}{4\pi^2 a^3} \int k^2 d k\,  \left[ \frac{1}{\omega_{\rm eff}}+ \frac{\mu^2}{2\omega_{\rm eff}^3} + \frac{(\frac16 -\xi)R}{2\omega_{\rm eff}^3} + \left(\frac{m_{\rm eff}^2\dot a^2}{2a^2\omega_{\rm eff}^5} + \frac{m_{\rm eff}^2\ddot a}{4a\omega_{\rm eff}^5} -\frac{5m_{\rm eff}^4\dot a^2}{8a^2\omega_{\rm eff}^7}\right)\right] \ . \ee
The last terms in the parentheses are finite and can be integrated to give
\be  \frac{1}{4\pi^2 a^3} \int k^2 d k\,  \left[ \frac{m_{\rm eff}^2\dot a^2}{2a^2\omega_{\rm eff}^5} + \frac{m_{\rm eff}^2\ddot a}{4a\omega_{\rm eff}^5} -\frac{5m_{\rm eff}^4\dot a^2}{8a^2\omega_{\rm eff}^7}\right] =  \frac{R}{288 \pi^2}
\ . \ee
Finally we obtain the same subtraction term as in the pragmatic mode-sum regularization method
\be\label{GeneralizedFinalCountertermAD}
^{(2)}G_{Ad}(x,x) =
\frac{1}{4 \pi^2 a^3}  \int_0^{\infty} d k k^2 \,  \left[ \frac{1}{\omega_{\rm eff}} + \frac{(\frac{1}{6} -\xi)R}{2\omega_{\rm eff}^3} + \frac{\mu^2}{2\omega_{\rm eff}^3}\right] + \frac{R}{288 \pi^2} \, .
\ee
Notice that this result agrees with \eqref{GeneralizedFinalCountertermPMS} in the coincidence limit, $\epsilon \rightarrow 0$, and both agree with \eqref{TPFfinal} when $\mu = 0$.

\section{Higher order expansion}\label{AppendixFourOrder}
In this appendix we expand the two-point function $G_{DS}(x,x')$ to order $\epsilon^2$. 
This expansion is enough to compute the vacuum expectation value of the stress-energy tensor $\langle T^{\mu\nu}\rangle$ by acting with a nonlocal operator to the symmetric part of the renormalized two-point function,  $\langle T_{\mu\nu}\rangle = \lim_{x \rightarrow x'}\mathcal{D}_{\mu\nu}(x,x')\left[\left\langle \{\phi(x), \phi\left(x^{\prime}\right)\}\right\rangle-G^{(1)}_{DS}(x,x') \right]$ \cite{bernard-folacci1986,christensen1976}. 

We begin by expanding \eqref{GreenFuncProperTime1} to linear order in $\sigma$ 
and up to and including four derivatives of the metric, which leads us to \eqref{christensen3.17}.
For simplicity we will deal with the case $\xi= \frac16$. 
The following expansions are enough to build the renormalized stress-energy tensor ($\sigma^\alpha \equiv \sigma^{;\alpha}$):
\bea
\Delta^{1 / 2}&&=1-\frac{1}{12} R_{\alpha \beta} \sigma^{\alpha} \sigma^{\beta}-\frac{1}{24} R_{\alpha \beta ; \gamma} \sigma^{\alpha} \sigma^{\beta} \sigma^{\gamma}+\left(\frac{1}{288} R_{\alpha \beta} R_{\gamma \delta}+\frac{1}{360} R^{\rho}{ }_{\alpha}{ }^{\tau}{ }_{\beta} R_{\rho \gamma \tau \delta}-\frac{1}{80} R_{\alpha \beta ; \gamma \delta}\right) \sigma^{\alpha} \sigma^{\beta} \sigma^{\gamma} \sigma^{\delta}+\cdots  \ , \ \ \ \ \label{delta12} \\
a_{1}&&=
\left[\frac{1}{90} R_{\alpha \rho} R^{\rho}{ }_{\beta}-\frac{1}{180} R^{\rho \tau} R_{\rho \alpha \tau \beta}-\frac{1}{180} R_{\rho \tau \kappa \alpha} R^{\rho \tau \kappa}{ }_{\beta} +\frac{1}{120} R_{\alpha \beta ;\rho}{ }^{\rho}-\frac{1}{360} R_{; \alpha \beta} \right] \sigma^{\alpha} \sigma^{\beta}+\cdots,\label{a1} \\
a_{2}&&=-\frac{1}{180} R^{\rho \tau} R_{\rho \tau}+\frac{1}{180} R^{\rho \tau \kappa \iota} R_{\rho \tau \kappa \iota}-\frac{1}{180} R_{; \rho}{ }^{\rho}+\cdots \ ,\label{a2}
\eea
where $\sigma^{\alpha}$ is computed up to order $\epsilon^5$ using the expansion (\ref{sigmaAll}). 
Expanding \eqref{christensen3.17} with \eqref{delta12}, \eqref{a1}, and \eqref{a2} we arrive at the following expansion for the two-point function up to and including the order $\mathcal{O}(\epsilon^2)$:
\bea
G^{(1)}_{DS}(x,x')&&=\frac{1}{4 \pi ^2 \epsilon ^2 a^2}+\frac{1}{480 \pi ^2 m^2 }\left[10 m^2 \left(\frac{ \ddot{a}}{a}+ \frac{ \dot{a}^2}{a^2}\right)-\left( \frac{a^{(4)}}{a}+\frac{\ddot{a}^2}{a^2}\right)+3 \left(\frac{\dot{a}^2\ddot{a}}{a^3} -
   \frac{ a^{(3)}}{a^2} \dot{a}\right)+60 m^4 \left(\gamma -\frac12 + \log \left(\frac{m \epsilon}{2}a \right) \right) \right]\nonumber\\
   &&+\frac{\epsilon ^2}{2880 \pi ^2} \left[\left(\frac32 a^{(4)}a +6 \ddot{a}^2 -\frac{\dot{a}^4}{a^2} +\frac{21}{2} a^{(3)} \dot{a}+\frac{23}{2} \frac{ \dot{a}^2 \ddot{a}}{a}\right)
  +30 m^2\left(a\ddot{a} +2\dot{a}^2\right)\left(\gamma -\frac12+ \log \left(\frac{m \epsilon}{2}a \right)\right)\right.\nonumber \\
  &&\left. +45 m^4 a^2\left(\gamma-\frac54+
   \log \left(\frac{m \epsilon}{2}a \right)\right) \right]+\mathcal{O}\left(\epsilon ^3\right) \ , \label{GDS4eps}
\eea
where $a^{(4)}\equiv \ddddot{a}$ and $a^{(3)}\equiv \dddot{a}$. This expression contains terms with four derivatives of the metric ($a^{(4)},\dot{a}^4,\dot{a}a^{(3)},\cdots$).

On the other hand, \eqref{GDS4eps} agrees with
\bea
\label{adiabaticST2sp} ^{(4)}G^{(1)}_{Ad}(x,x')  &=&\frac{1}{4\pi^2 a^3}  \int_0^\infty k^2dk \frac{\sin k\epsilon}{k\epsilon} \left[ \frac{1}{\omega} + \frac{(\frac{1}{6} -\xi)R}{2\omega^3} +\frac{m^2\dot a^2}{2a^2\omega^5} + \frac{m^2\ddot a}{4a\omega^5} -\frac{5m^4\dot a^2}{8a^2\omega^7}+(W^{-1})^{(4)}\right] \ ,
\eea 
when it is expanded at order $\mathcal{O}(\epsilon^2)$. Equation \eqref{adiabaticST2sp} is the expansion of adiabatic regularization at fourth adiabatic order \cite{parker-toms}. The integral on $(W^{-1})^{(4)}$ is finite and contains terms with four derivatives of the metric.

\end{document}